\documentclass[rnote]{aa}
\usepackage{graphicx}
\usepackage{natbib}
\usepackage{txfonts}
\begin{document}

\def\qj{QJ~0158-4325}
\def\kmsmpc{~km~s$^{-1}$~Mpc$^{-1}$}
\def\ho{H$_0$}
\def\ctq{QJ~0158-4325}
\def\zl{z$_{\rm lens}~$}
\def\kms{km s$^{-1}$}

 \title{Redshifts and lens profile for the double quasar  \qj \thanks{Based on observations made with ESO Telescopes at Paranal Observatory programme ID: 67.A-0502(A,B) and with  NASA/ESA Hubble Space Telescope, obtained from the data archive at the Space Telescope Institute. STScI is operated by the association of Universities for Research in Astronomy, Inc. under the NASA contract  NAS 5-26555.}}

  \subtitle{}
   
  \author{C. Faure\inst{1,2}  \and T. Anguita\inst{2}   \and A. Eigenbrod\inst{1}  \and J.-P. Kneib\inst{3}\and V. Chantry\thanks{Research Fellow, Belgian National Fund for Scientific Research (FNRS)}\inst{4}  \and D. Alloin\inst{5} \and N. Morgan\inst{6}   \and G. Covone\inst{7}}
\institute{Laboratoire d'Astrophysique, Ecole Polytechnique F\'ed\'erale de Lausanne (EPFL), Observatoire de Sauverny, 1290 Versoix, Switzerland
\and
Astronomisches Rechen-Institut, Zentrum f\"{u}r Astronomie der Universit\"{a}t Heidelberg, M\"{o}nchhofstr. 12-14, D-69120, Heidelberg, Germany	
        \and
Laboratoire d'Astrophysique de Marseille, UMR6110, CNRS-Universit\'e de Provence, 38 rue Fr\'ed\'eric Joliot-Curie, F-13388 Marseille Cedex 13, France 
	\and 
Institut d'Astrophysique et de G\'eophysique, Universit\'e de Li\`ege, All\'ee du 6 ao\^ut, 17, Sart Tilman (Bat. B5C), Li\`ege 1, Belgium
	\and
Laboratoire AIM, CEA/DSM-CNRS-Universit\'e Paris Diderot, IRFU/Service d'Astrophysique, B\^at. 709, L'Orme des Merisiers, CEA Saclay, F-91191 Gif sur Yvette C\'edex, France
	\and
Department of Astronomy, The Ohio State University, 140 W. 18th Avenue, Columbus, OH 43210-1173, USA
	\and
Dipartimento di Scienze Fisiche, Universit\'a di Napoli ``Federico II'', Naples, Italy }

\date{Accepted  05/12/2008}

\authorrunning{C. Faure et al.}
\titlerunning{Redshifts and lens profile for QJ~0158-4325}

\abstract{}{We report on the redshift of the lensing galaxy and of the quasar \qj\, and  on the lens model of the system.}{A deep VLT/FORS2 spectrum  and  HST/NICMOS-F160W   images  are   deconvolved.  From the images we derive the light profile of the lensing galaxy and an accurate relative astrometry for the system.  In addition we measure the flux ratio between the quasar images in the \ion{Mg}{II} emission line to constrain the mass model.}{ From the spectrum we measure the redshift of the lensing galaxy ($z=0.317\pm 0.001$) and of  the quasar ($z=1.294\pm0.008$). Using the flux ratio in the lens model allows to discard the SIE as a suitable  approximation of the lens potential. On the contrary the truncated-PIEMD  gives a good fit to the lens and leads to a time delay of $\Delta$t$_{A-B}$=-14.5$\pm$0.1~days, with \ho=73\kmsmpc.}{Using the flux ratio to constrain the mass model favors the truncated-PIEMD over the SIE, while ignoring this constraint leaves the choice open. }{}

 \maketitle
\section{Introduction}\label{intro}
The study of  gravitationally lensed quasars is relevant both to cosmology and  to the determination of the total mass in (lensing) galaxies. Time delays between multiple quasar images depend both on the mass distribution of the lensing galaxy and on the Hubble parameter, H$_0$ (Refsdal, 1964). The translation of an observed time delay into a value of \ho\, requires the modelling of the surface mass density of the lens.  For that purpose, geometric and photometric constraints are needed such as: (a) spectroscopic redshifts for both the source and the lens components and (b) accurate relative positions of the lensed images and of the lensing galaxy. Additional constraints to be used are the flux ratios (FRs) between the quasar images. Indeed, the FRs are equal to the magnification ratios at the image positions, and hence depend on the mass distribution of the lens. However, the contribution of the lens (lensing galaxy and environment) must be disentangled from other sources of flux variation such as:  differential galactic extinction, intrinsic quasar variability, microlensing and/or millilensing (de)magnification, all phenomena which have different time scales and wavelength dependencies (see e.g. Anguita et al. 2008 for more details). In the case of radio loud quasars this is rather straightforward, yet it does not apply to the whole quasar population. For optically selected quasar, the separation of the different sources of flux variation is more complex. In spite of this complexity, one must attempt it to derive values of the image FRs genuinely related to the strong lens. In the case of doubly imaged quasars (which constitute $\sim$60\% of the strongly lensed quasars known so far), it is particularly important to use the FR to constraint the lens mass model, as these systems are  naturally the least constrained. \\
In this paper we report on new observations of  the lensed quasar \qj, and on a model of the lensing system. The quasar and the  datasets relevant to our study are presented in $\S$~\ref{theqso}. In $\S$~\ref{deconvsec}, the spectra and image deconvolutions are discussed. We model the lens in $\S$~\ref{simple}. Conclusions are given in $\S$~\ref{dandc}.  \\
We assume a WMAP type cosmology (Spergel et al. 2003, 2007): $H_0$=73\kmsmpc, $\Omega_m$=0.3 and $\Omega_\Lambda$=0.7. All magnitudes quoted in the paper are in the AB system.
\section{\qj\, and dataset}\label{theqso}
 The quasar was detected in the Cal\'an-Tololo Quasar survey (Maza et al. 1995), therefore called CTQ~414, and confirmed to be a lensed quasar (at z=1.29) by  Morgan et al. (1999). The two images of the quasar are separated by 1.22\arcsec. A galaxy overdensity was detected along the line-of-sight to the quasar (Faure et al. 2004), suggesting the presence of a galaxy group. \\
 There  remains in the literature a discrepancy of $\sim$0.8\arcsec\, between the absolute declination of image A (see label in Fig.~\ref{deconv}) by Morgan et al (1999) and that provided in the CASTLES database (Mun\~oz et al 1998). In order to elucidate this offset, we have matched the USNO-B1 stellar catalogue to the FORS R-band  images of the lens (presented in Faure et al. 2004): the RMS values are 0.07\arcsec\, and 0.06\arcsec\, in RA and DEC  respectively. We have then derived the absolute astrometry of image A: $\alpha_{J2000}=$01$^{h}$58$^{m}$41.43$^{s}$ and $\delta_{J2000}=$-43$^{o}$ 25\arcmin 3.5\arcsec. \\
Recently Morgan et al. (2008) have  published the continuum light curves of the quasar images, monitored during four seasons at the Apache Point Observatory (SMARTS) and at ESO/La Silla (Euler Swiss Telescope). However, they were not able to derive a reliable time delay, as microlensing events overwhelmed the intrinsic variability, and their best guess, for a realistic mass model (image A leading) is: $\Delta t_{AB}\sim$-20 to -10~days. The system is still under photometric  monitoring.\\
We have retrieved NICMOS-2 (Near Infrared Camera and Multi-Object Spectrometer)  F160W  images of \qj\, from the HST archival data related to a large public survey of gravitational lenses (PI: Falco, CASTLES\footnote{http://cfa-www.harvard.edu/castles}). \\
From the VLT archive, we have retrieved a set of four long-slit spectra centered on the quasar images and on the lensing galaxy, obtained with FORS2 (PI: Rix).
Each spectrum has an exposure time of 1300~s. The spectra were taken through an 0.7\arcsec slit, under a mean seeing of 1.28\arcsec. The resolution is 4.86~\AA\,pix$^{-1}$, from 5500~\AA\, to 11000~\AA. 
The spectra have been reduced following Eigenbrod et al. (2006). It is not possible to disentangle the quasar light from the light of the lensing galaxy through a direct analysis of the spectra. However, the spectra are suitable for spectral deconvolution. \\
\section{Deconvolution}\label{deconvsec}
\subsection{Spectral analysis and flux ratio} \label{decspeceg}
\begin{figure}[h!]
\begin{center}
\includegraphics[width=8.7cm]{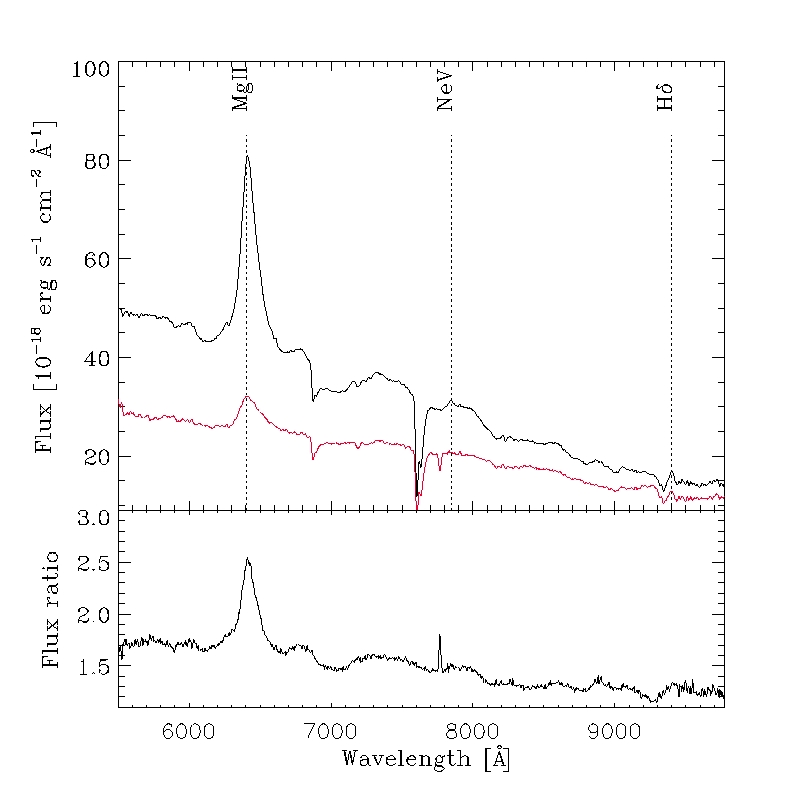}
\caption{Spectra of the images of \qj. Top spectrum:  image A. Bottom spectrum:  image B. Bottom panel: ratio between the two spectra.}
\label{Q0158_qso}
\end{center}
\end{figure}

\begin{figure}[h!]
\begin{center}
\includegraphics[width=8.7cm]{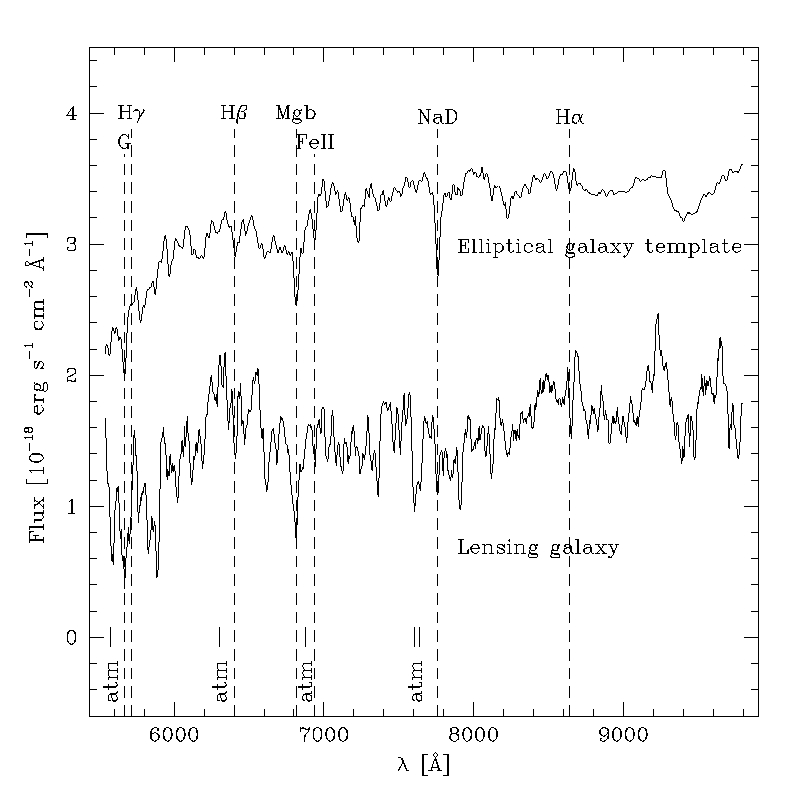}
\caption{Spectrum of the lensing galaxy. The total integration time is 5000~s. The template spectrum of a redshifted 
elliptical galaxy at z=0.317 is shown for comparison (Kinney et al. 1996).
Atmospheric absorptions are indicated by the  label {\tt atm}.
}
\label{Q0158_lens}
\end{center}
\end{figure}

In order to measure the redshift of the lensing galaxy  we have applied the spectral version of the MCS deconvolution algorithm (Magain, Courbin \& Sohy 1998, Courbin 1999) to separate the  lensing galaxy from the  much brighter quasar images in the FORS2 spectroscopic dataset. The long-slit FORS2 data do not sample any star. To circumvent this difficulty, we have modified the deconvolution strategy, compared to previous studies (e.g. Eigenbrod et al. 2008): we do not use stars, but instead the two bright quasar images as point spread function (PSF) references. We then deconvolve the data considering three point sources (two for the quasar images A and B, and one for the lensing galaxy). Along the deconvolution, the software optimizes the positions and intensities of the spectra, and at the end yields the individual spectra of the three point sources. The spectra of the quasar images  display strong \ion{Mg}{II} and weak [NeV] and H$\delta$ emission lines (Fig.~\ref{Q0158_qso}). The bright \ion{Mg}{II} doublet is used to derive the redshift, through a two-line fit. Using the measured line centers and the rest wavelengths [2795.5\AA{}, 2802.7\AA{}] for the \ion{Mg}{II} doublet, we deduce a redshift z=1.294$\pm$0.008 for the quasar.  Fig.~\ref{Q0158_lens} shows the final one-dimensional spectrum of the lensing galaxy, smoothed with a 24~\AA\ box. Residuals of the quasar \ion{Mg}{II} line are negligible, assessing the efficiency of the spectral decomposition. The lensing galaxy spectrum exhibits the typical absorption lines of an elliptical galaxy at z=0.317$\pm$0.001 (based on the position of the following lines: Gband, H$\beta$, Mgb, FeII, NaD and H$\alpha$). 

In $\S$~\ref{intro} we have listed the phenomena which can affect the observed FRs between quasar images.  Let us discuss now the FR genuinely due to the lens and the confidence with which it can be used in the lens mass model. \\
The quasar image spectra exhibit clearly  the \ion{Mg}{II} line, which is expected  to arise from a region in the quasar, (i) which is too large to be sensibly magnified by microlensing (Abajas et al. 2002, Eigenbrod et al. 2008), and (ii) with intrinsic flux variation appreciable over a time scale  of months to years (Peterson 2001). Therefore, this broad, low ionization emission line is mainly affected by galactic extinction and by the lens magnification. Another source of flux variation could be substructures associated to the lensing galaxy on the milli-arcsecond scale (such as a dark matter clump or a dwarf galaxy). In the available dataset, we have no evidence for the presence of such substructures. Therefore we assume that there is no perturbation of such scale.\\
The FR in the \ion{Mg}{II} line  is a function  of the flux measured in the emission line, $E_i$, and in the continuum, $C_i$. The ratio between images A and B spectra (Fig.~\ref{Q0158_qso}, bottom panel) exhibits two remarkable facts, (i) a different value of the FR depending on whether it is measured in the continuum or in the \ion{Mg}{II} line and (ii) a slightly negative slope. 
As to the first fact, the quasar continuum is associated with the emission of an accretion disk, the projected size of which is comparable to the Einstein radii of stars in the lensing galaxy (e.g. Anguita et al 2008). Therefore, the continuum emission is prone to be magnified (or de-magnified) by individual stars in the lensing galaxy. The intrinsic variation of this region may happens to be fast (days or weeks).  As to the second fact, the slope in the continuum FR could be explained either by microlensing or by galactic extinction. \\
The continuum and emission line ratios  vary for different reasons, hence to  measure properly the FR in the emission line it is necessary to remove the contribution of the continuum. The FR in the \ion{Mg}{II} line is given by: $\frac{F_A}{F_B}$=$\frac{E_A-C_A}{E_B-C_B}$=5.52 (or $\Delta(m_B-m_A)$=1.85~[mag]). The quasar image B being closest to the lensing galaxy center, it is at the same time less magnified by the lens and more extinguished by the lensing galaxy, than image A. Hence, the value $\Delta(m_B-m_A)$=1.85~[mag] is an upper limit of the difference due to the potential of the lens. Yet, the elliptical nature of the lensing galaxy tells us that the differential galactic extinction is likely to be fairly small and that the lens magnification is indeed the dominant contributor.

\subsection{Image deconvolution and  galaxy light profile}\label{imdec}

\begin{figure*}
\begin{center}
\includegraphics[width=12cm]{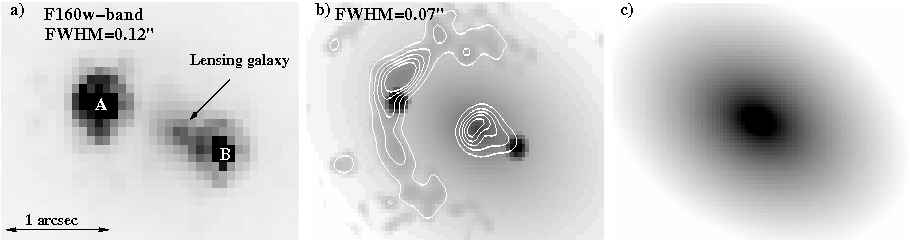}
\caption{\label{deconv} Deconvolution of \qj\ images (F160W). North is to the top, East to the left. {\it a}) Original image.  {\it b}) Deconvolved image: the quasar images A and B appear as point sources, the lensing galaxy is the sum of a de Vaucouleurs profile plus the residuals which are not fitted by the light profile. The white contours trace the light distribution of the smooth objects (quasar host galaxy and flux residuals of the lensing galaxy). (c) de Vaucouleurs luminosity profile of the lensing galaxy, with a lower flux cut than in image (b), in order to highlight its orientation and ellipticity. }
\end{center}
\end{figure*}

Moreover, we have applied the MCS deconvolution code to the  NICMOS/F160W images. A technique to build the PSF when no star is available in the field is the iterative method described in Chantry \& Magain (2007).
  As a starting point, we used {\it Tiny Tim} PSFs (Krist \& Hook 1997) already improved by the iterative method applied on another lensed system (the Cloverleaf, see Chantry \& Magain 2007) observed under similar conditions. Then we deconvolve simultaneously the four NICMOS exposures.  Doing so we obtain a good estimate of the background level, which we subtract from the initial individual exposures. The images cleaned from their background are then deconvolved simultaneously to reach a final FWHM=0.07\arcsec. The result is displayed in Fig.~\ref{deconv}. The quasar host galaxy is visible under quasar image A.\\
In addition, a fit of the lensing galaxy luminosity profile is performed during the deconvolution. A good fit is obtained with the de Vaucouleurs profile, as already mentioned in Rusin et al. (2003) and Morgan et al. (2008). We find the following parameters for the lensing galaxy light profile: PA =61$^\circ$, $\epsilon$=0.32$\pm$0.01, r$_{eff}$=0.70$\pm$0.01\arcsec. The magnitude and relative position are displayed in Table~\ref{photom}.  The value of the effective radius is in agreement with a previous study of this system by Rusin et al. (2003, r$_{eff}$=0.66$\pm$0.04\arcsec) for the lensing galaxy, using data from the HST WFPC (F814W and F555W) and the same NICMOS (F160W) observations.  In the band F160W, they found the lensing galaxy to be $\sim$0.2~mag brighter than in the present study: an explanation would be that the light of the host galaxy under the quasar image A was not removed properly, leading to an overestimation of the lensing galaxy total flux.  In Table \ref{photom},  the error bars  correspond to the standard deviations of the mean values measured in the individual deconvolved frames.

\begin{table}
\renewcommand{\arraystretch}{1.0}
\centering
\begin{center}
\caption{\label{photom} Summary of astrometric, photometric and redshift measures.}
\begin{tabular}{l l l l l l l }
\hline
\hline
            & A &B &G     \\
\hline
$\Delta \alpha$ ('') &  0 & -1.156$\pm$0.001 &  -0.780$\pm$0.001 \\%

$\Delta \delta$ ('') &  0 &-0.398$\pm$0.002 &  -0.246$\pm$0.002 \\%
\hline
z&  1.294$\pm$0.008 & 1.294$\pm$0.008 &  0.317$\pm$0.001 \\
\hline 
F160W &  16.24 $\pm$0.02 & 17.02$\pm$0.02 &   16.88$\pm$0.05  \\
\hline
\end{tabular}
\end{center}
\end{table}



\section{Mass models}\label{simple}

We use the Lenstool code (Kneib et al. 1993, Jullo et al. 2007) to model the lens, where the  singular isothermal ellipsoid (SIE) is parameterized by its position, its  position angle, its ellipticity and its velocity dispersion.  During the modelling, the position, ellipticity and position angle of the lens are fixed to the values measured on the deconvolved NICMOS images (see $\S$~\ref{imdec}). The redshifts of the source and lensing galaxy are fixed to the new values measured in this paper. The only free parameter is the velocity dispersion. We find an acceptable fit ($\chi^2\sim$18) for the SIE, but the predicted FR between the quasar images is four times larger than observed (see $\S$~\ref{decspeceg}). The predicted time delay is -12.2$\pm$0.1 days. If we force the FR to be equal to the value measured in the \ion{Mg}{II} line, and in exchange let free the position angle of the galaxy or its ellipticity, we do not find any good fit ($\chi^2>>$100), showing that a single SIE cannot explain such a low FR between the quasar images. \\
Then, we model the lens with a truncated pseudo-isothermal elliptical  mass distribution (TPIEMD; Kassiola \& Kovner 1993, Kneib et al. 1996) for both the bright and dark components of the lens.  The TPIEMD has a  core radius $r_{core}$ and cut radius r$_{cut}$: we fix $r_{cut}$=$\frac{4}{3}$r$_{eff}$ (see Eliasdottir et al. 2007) and $r_{core}$=0.01\arcsec, with the aim of keeping  r$_{core} << $r$_{cut}$. By fixing r$_{cut}$ as a function of r$_{eff}$ we effectively define a constant mass to light ratio lens potential. Doing so, the number of free parameters is reduced to one: the velocity dispersion. The FR predicted  is slightly larger than the upper limit calculated from the spectra. If we force the FR to match the observed one and free the position angle or the ellipticity (within the error bars given by the deconvolution) we obtain a good fit ($\chi^2$=2.15). Using a different value for  r$_{core}$  (equal to 0.005\arcsec\, or  0.05\arcsec)  modifies slightly the time delay (respectively -14.2$\pm$0.1~days and  -14.7$\pm$0.1~ days). Results of the models are summarized in Table~\ref{modelsimple}. The error bars correspond to $\Delta \chi^2$=1. \\

\begin{table}
\renewcommand{\arraystretch}{1.0}
\centering
\caption{Parameters of  the  lens mass models. 
Col. 2: $\chi^2$ of the fit.
Col. 3: Velocity dispersion (km~s$^{-1}$).
Col. 4: Time delay  in days (image A leading).
Col. 5: FR  (image A brighter).}
\label{modelsimple}
\begin{tabular}{c| c c c c c }
\hline
\hline
 & $\chi^2$&$\sigma_v$      & TD& FR  \\
\hline
SIE & 18.8  & 183.5$\pm$0.2         &-12.1$\pm$0.1 &23.4$\pm$1.3 \\
TPIEMD &3.8 & 179.9$\pm$0.2 &-14.4$\pm$0.1 &6.5$\pm$0.3\\
TPIEMD &4.7 &179.7$\pm$0.2 &-14.5$\pm$0.1 &5.52 (fixed)\\
\hline
\hline
\end{tabular}
\end{table}


\section{Conclusions}\label{dandc}
 We have measured the redshift of the lensing galaxy and improved the measurement of the quasar redshift in \qj. While modelling the lens, we noticed that using only the quasar image positions to constrain the mass model does not allow to discriminate a mass density profile over another: in the case of \qj, both a SIE and a TPIEMD appear as suitable approximations of the lensing potential. If we add the constraint of the quasar image FR (measured in the \ion{Mg}{II} line), it becomes possible to discard the SIE as a suitable approximation of the lensing potential. The lens modelled as a TPIEMD is good. The time delay inferred with the TPIEMD is $\Delta$t$_{A-B}$=-14.5$\pm$0.1 days for $H_0$=73\kmsmpc.\\
In conclusion, using the  FR to constrain the lens allows to select the most suitable galaxy profile and to measure its mass, as well as to predict the time delay with more confidence (or the Hubble constant if the time delay is known).  

\begin{acknowledgements} We thanks F. Courbin for discussions and suggestions. C.F. thanks G. Letawe for giving access to the new version of the MCS software. This work started under  the support from the European Community's Sixth Framework Marie Curie Research Training Network Programme, Contract No.  MRTN-CT-2004-505183 ``ANGLES''. D.A. acknowledges travel support from ESO, CNRS and CEA at different stages of advancement of this paper. J.-P.K. acknowledges support from CNRS. The HST data used  in this paper were obtained by the ``CfA  Arizona Space Telescope  LEns Survey''  (CASTLES) collaboration (PI:  E.  Falco)  
\end{acknowledgements}

\end{document}